\documentstyle[12pt,twoside,cosmion,epsf,rotate]{article}
\heads{High Energy Astrophysics: Accretion Theory}
{Neutron stars in X-ray pulsars and their magnetic fields \ \ {\rm S.B. Popov
and V.M. Lipunov}}
\begin{document}

\Arthead{1}{1}

\Title{Neutron stars in X-ray pulsars and their magnetic fields}
{S.B. Popov$^1$, V.M. Lipunov$^{1,2}$}
{$^1$Sternberg Astronomical Institute, Universitetskij pr. 13, 119899 Moscow
\\
$^2$Physics Department, Moscow State University}

\Abstract{
 Estimates of the magnetic field of neutron stars in X-ray pulsars
are obtained using the hypothesis of the equilibrium period for disk and
wind accretion and also
from the BATSE data on timing of X-ray pulsars
using the observed maximum spin-down rate.
Cyclotron lines at energies $\ge 100$ keV in several Be-transient are
predicted for future observations.

 We suggest a new method of estimating distances to X-ray pulsars
and their magnetic fields.
Using observations of fluxes and period variations in the model
of disk accretion one can estimate the magnetic momentum of a neutron star
and the distance to X-ray pulsar.
 As an illustration the
method is applied to the system GROJ1008-57.}

\section{Introduction}

Among all astrophysical objects neutron stars (NSs)
attract most attention of physicists.
Now we know more than 1000 NSs as 
radiopulsars and more than 100 NSs emitting X-rays,
but the Galactic population of these objects is about $10^8$ -- $10^9$.
So only a tiny fraction of one of the most fascinating astrophysical objects
is observed at present. 

 NSs can appear as sources of different nature:
as isolated objects 
and as binary companions, 
powered by wind or disk
accretion from a secondary companion.
X-ray pulsars are probably one of 
the most prominent among binary sources, because there important
parameters of NSs can be determined.

 Now we know more than 40 X-ray pulsars 
(see e.g. Bildsten et al.\cite{1}, Borkus\cite{2}). 
Observations of optical
counterparts of X-ray sources
give an opportunity to determine distances to these objects 
and other parameters with relatively
high precision, and with hyroline detections one can obtain the value of
magnetic field, $B$, 
of a NS. But lines are not detected in all sources of that
type 
and magnetic field can be estimated
from period measurements (see e.g. Lipunov\cite{4}). Precise distance
measurements usually are not available immediately after X-ray discovery
(especially, if localization
error boxes are large and X-ray sources have transient nature).
In that sense methods of simultaneous determination of field and distance
basing only on X-ray observations can be useful, and several of them were
suggested by different authors previously.

 Here we try to obtain estimates of the magnetic fields (and distances)
of NSs in X-ray pulsars from their period (and flux) variations.

\section{Estimates of the magnetic field}

 Magnetic fields of accreting NSs can be estimated using period
variations or using the hypothesis of the equilibrium period (see
Lipunov\cite{4}). We use both of these methods.

 For estimating of magnetic momentum of NSs using observed values
of maximum spin-down we use the following main equation:

$$
\frac{dI\omega}{dt}=-k_t\frac{\mu^2}{R_{co}^3},
$$
where $I$ -- NS's momentum of inertia, $\omega=\frac{2\pi}p$ 
-- spin frequency, $\mu$ -- magnetic momentum, $R_{co}=\left
(\frac{GM}{\omega^2}\right)^{1/3}$-- corotation radius. 
We used $k_t=1/3$, $I=10^{45}$ g cm$^2$, $M=1.4M_{\odot}$.

We used graphs from (Bildsten et al.\cite{1}) to derive spin-up
and spin-down rates and flux changes measurements.  Data on these graphs is  
shown with one day time resolution. 

 Equilibrium period can be written in different forms for disk and wind-fed
systems. For the first case we used the following equation:

\begin{equation}
p_{eq. disk}=2.7\, \mu_{30}^{6/7} L_{37}^{-3/7}\, s.
\end{equation}

For wind-accreting systems we have:

\begin{equation}
p_{eq. wind}=10.4\,L_{37}^{-1}T_{10}^{-1/6}\mu_{30} \, s.
\end{equation}
Here $L_{37}$ -- luminosity in units $10^{37}$ erg s$^{-1}$,
$T_{10}$ -- orbital period in units 10 days, $\mu_{30}$ -- magnetic momentum
in units $10^{30}$ G cm$^{3}$. 

 Estimates of the magnetic momentum, $\mu$, obtained with different
assumptions are shown in the table 1.
Three values are shown: an estimate from spin-down obtained from the BATSE
data (Bildsten et al.\cite{1}); 
an estimate from the equilibrium period for wind-fed systems (eq. (2));
an estimate for disk-accreting systems (eq. (1)). 
Less probable values 
are marked with asterix. 

\begin{table}[h]
\caption[]{Spin-down and magnetic momentum estimates}
\begin{tabular}{|l||c|c|c|c|c|}
\hline
X-RAY & maximum & Source & Magnetic & Magnetic & Magnetic \\
 PULSAR & dp/dt   & Type & momentum & momentum &momentum \\
        & observ. &     & (spin-down), & (wind), &(disk),\\
        & (spin-down)           &     & $10^{30}$ G cm$^{3}$ &$10^{30}$ G cm$^{3}$ & 
$10^{30}$ G cm$^{3}$ \\
\hline
 GRO 1744-28    &         &   BeTR   &         &    0.93$^*$  &  0.58\\
 HER X-1     & $9.3\cdot10^{-13}$ &   LMXRB  &     0.3 &          &  0.18\\
 4U 0115+63  & $3.0\cdot10^{-10}$ &   BeTR   &    5.17 &    0.32$^*$  &  1.26\\
 CEN X-3     & $7.5\cdot10^{-12}$ &   HMSG   &   0.82  &    1.8   &  4.42\\
 4U 1627-67  & $4.1\cdot10^{-11}$ &   LMXBR  &   1.9   &          &  2.82\\
 2S 1417-624 &         &   BeTR     &         &    8.64$^*$  &  17.82\\
 GRO 1948+32 & $5.4\cdot10^{-9}$  &   BeTR   &   22.0  &          &      \\
 OAO 1657-415& $1.5\cdot10^{-7}$  &   HMSG   &   115.1 &    0.15  &  4.33\\
 EXO 2030+375&         &   BeTR   &         &    0.1   &  3.45\\
 GRO 1008-57 & $3.2\cdot10^{-8}$  &   BeTR   &   53.3  &          &      \\
 A 0535+26   &         &   BeTR   &         &    30.24$^*$ &  101.23\\
 GX 1+4      & $6.4\cdot10^{-8}$  &   LMXRB  &   75.5  &          & 167.3\\
 VELA X-1    & $3.8\cdot10^{-9}$  &   HMSG   &   18.5  &    4.03  &  88.15\\
 4U 1145-61  & $3.3\cdot10^{-7}$  &   BeTR   &   172.1 &    0.23$^*$  &  16.7\\
 A 1118-616  & $5.1\cdot10^{-7}$  &   BeTR   &   212.8 &          &  245.5\\
 4U 1535-52  & $3.6\cdot10^{-7}$  &   HMSG   &   56.4  &    17.37 &  299.3\\
 GX 301-2    & $8.9\cdot10^{-6}$  &   HMSG   &   281.9 &    8.34  &  200.5\\
\hline
\end{tabular}
\end{table}

In table 1 we use the following
notation: LMXRB- Low Mass X-Ray Binary;
HMSG - High Mass SuperGiant;
BeTR- Be-transient source.
Values, which
were used for estimates with the hypothesis of
the equilibrium period: spin period,
mean luminosity in units $10^{37}$ erg s$^{-1}$, orbital period in units 10
days can be found on the Web:
{\bf http://xray.sai.msu.ru/\~\,polar}.

 More precise estimates can be made by fitting all observed values
of spin-up and spin-down rate together with flux measurements. 
When the distance to the source is know
only the value of the magnetic field should be fitted.
And on the figure 1 we show such estimates for Her X-1.

\begin{figure}
\epsfxsize=0.9\hsize
\centerline{\rotate[r]{\epsfbox{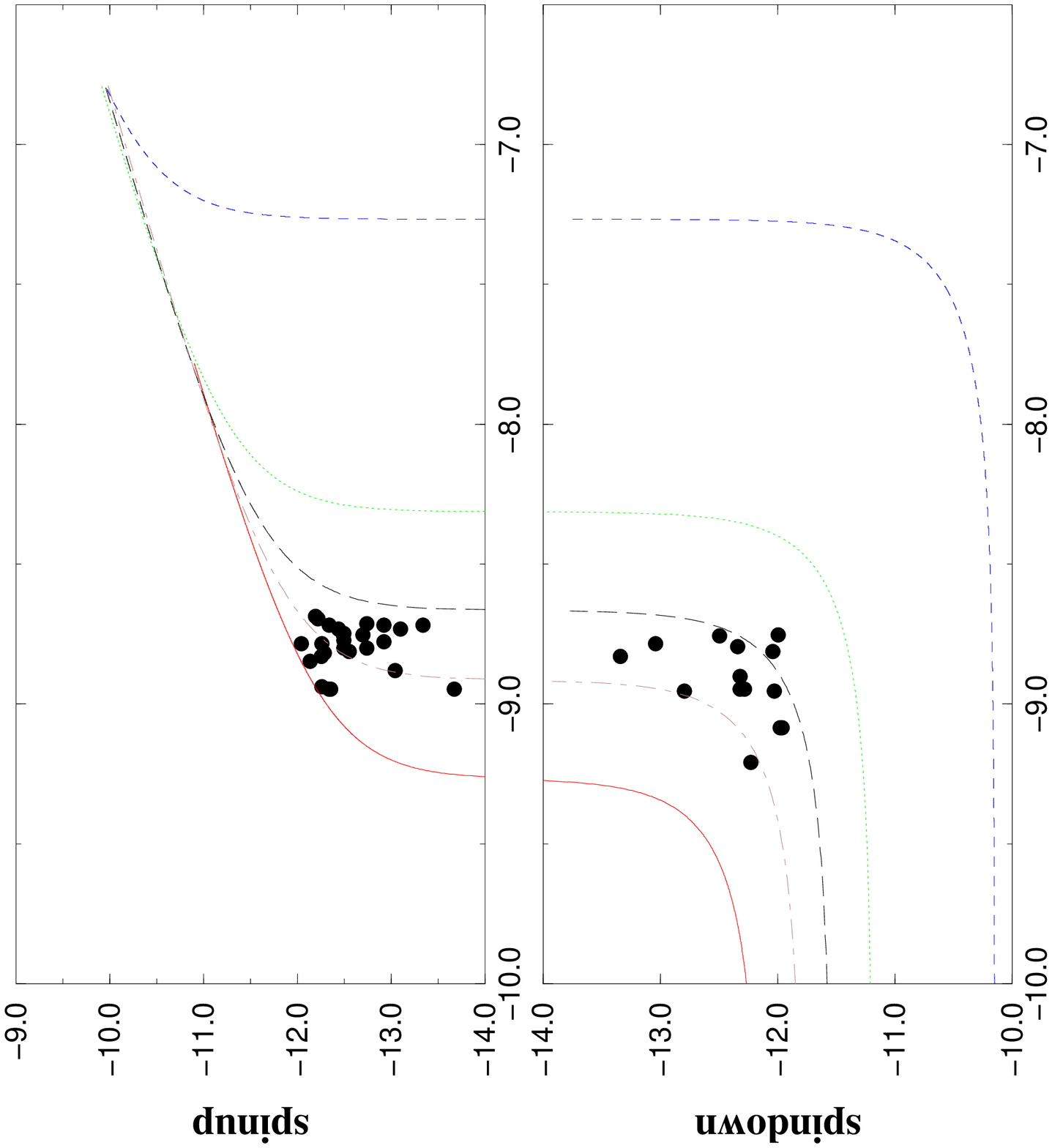}}}
\caption{Dependence of period derivative, $\dot p $, on the parameter $p^{7/3}f$,
$f$-- observed flux, for Her X-1. Both axis are in logarithmic scale.
Observations (Bildsten et al., 1997) are shown with black dots.
Five curves are plotted for different values of the magnetic field.
 Solid curve: $\mu=0.1\cdot 10^{30}\, {\rm G}\cdot {\rm cm}^3$.
Dot-dashed curve: $\mu=0.15\cdot 10^{30}\, {\rm G}\cdot {\rm cm}^3$.
Long dashed curve: $\mu=0.2\cdot 10^{30}\, {\rm G}\cdot {\rm cm}^3$.
Dotted curve: $\mu=0.3\cdot 10^{30}\, {\rm G}\cdot {\rm cm}^3$.
Dashed curve: $\mu=1\cdot 10^{30}\, {\rm G}\cdot {\rm cm}^3$.
All curves are plotted for the distance $d=4 \, {\rm kpc}$.}
\end{figure}

 We plot spin-up and spin-down rates as a function of the parameter,
which is a combination of the spin period and source's luminosity.
Spin-up and spin-down values derived
from the BATSE data (Bildsten et al.\cite{1})
are plotted as black dots,
and theoretical curves for different values of the magnetic momentum
are also shown. In ideal the best curve for the magnetic momentum should
exist, which fits all observational points. In reality points have some
errors, distance to the source in also know with some uncertainty, and
simple model of spin-up and spin-down can be only the first approximation.

\section{Discussion and conclusions}

 We made estimates of the magnetic field of NSs in X-ray pulsars.
Estimates which were 
made with an assumption that $p=p_{eq}$ are rather rough.
Obtained values depend (except uncertainties connected with the method
itself) on unknown parameters of NSs, such as masses, radii, moments
of inertia. All of them were accepted to have ``standard'' values,
and of course it is only the first approximation.
For example, our estimate for the source GRO 1744-28 is $\mu \sim 10^{30}$
G cm$^3$, and it is smaller than the estimate shown in (Borkus\cite{2}),
which is $B\sim (2-5)\cdot 10^{12}$ G (we mark, that the estimate obtained by
Joss \& Rappaport\cite{3} is significantly
lower than both: Borkus and our estimates). 
But if one take ``non-standard'' value for
$R$, these estimates of $\mu$ and $B$ can be in good correspondence.

We show several examples in table 2. NSs radii are calculated from the
following simple formula:

$$
R=\left(2\mu/B\right)^{1/3}.
$$
Here $\mu$ are taken from table 1, and values of $B$ are taken from
Nagase\cite{5}, Borkus\cite{2} and Wang\cite{8}. As one can see from the table
for several sources
measured $B$ are not in correspondence with our calculated $\mu$,
and radii of NSs are too big.
Mostly these cases are long period wind-fed pulsars like GX 301-2, 
where formation of temporal
reverse disk is possible for the cases of fast spin-down, 
so there maximum spin-down can be not the best field
estimate, and estimates from the equilibrium period for wind-accretion case 
are in better correspondence with observations.

\begin{table}[h]
\caption[]{Magnetic fields, magnetic momentum and radii}
\begin{tabular}{|l||c|c|c|}
\hline
X-RAY   & Magnetic &    Magnetic &   Neutron \\
 PULSAR & momentum &    field    &   star    \\
        & (calcul.), &  (observ.), &   radius,  \\
        & $10^{30}$G cm$^{3}$ & $10^{12}$G  & km \\
\hline
 GRO 1744-28    & 0.58 & $ \sim (2-5)$& $\sim (8.3-6.1)$ \\
 HER X-1     & 0.3  & 3& 5.8 \\
 4U 0115+63  & 5.17 & 1.1& 21.1 \\
 A 0535+26   & 101.23 &11& 26.4 \\
 VELA X-1    & 18.5 & 2.3 & 25.2 \\
 4U 1535-52  & 56.4 & 1.9 & 39 \\
 GX 301-2    & 281.9 & 3.5 & 54.4 \\
\hline
\end{tabular}
\end{table}

\begin{figure}
\epsfxsize=0.9\hsize
\centerline{\rotate[r]{\epsfbox{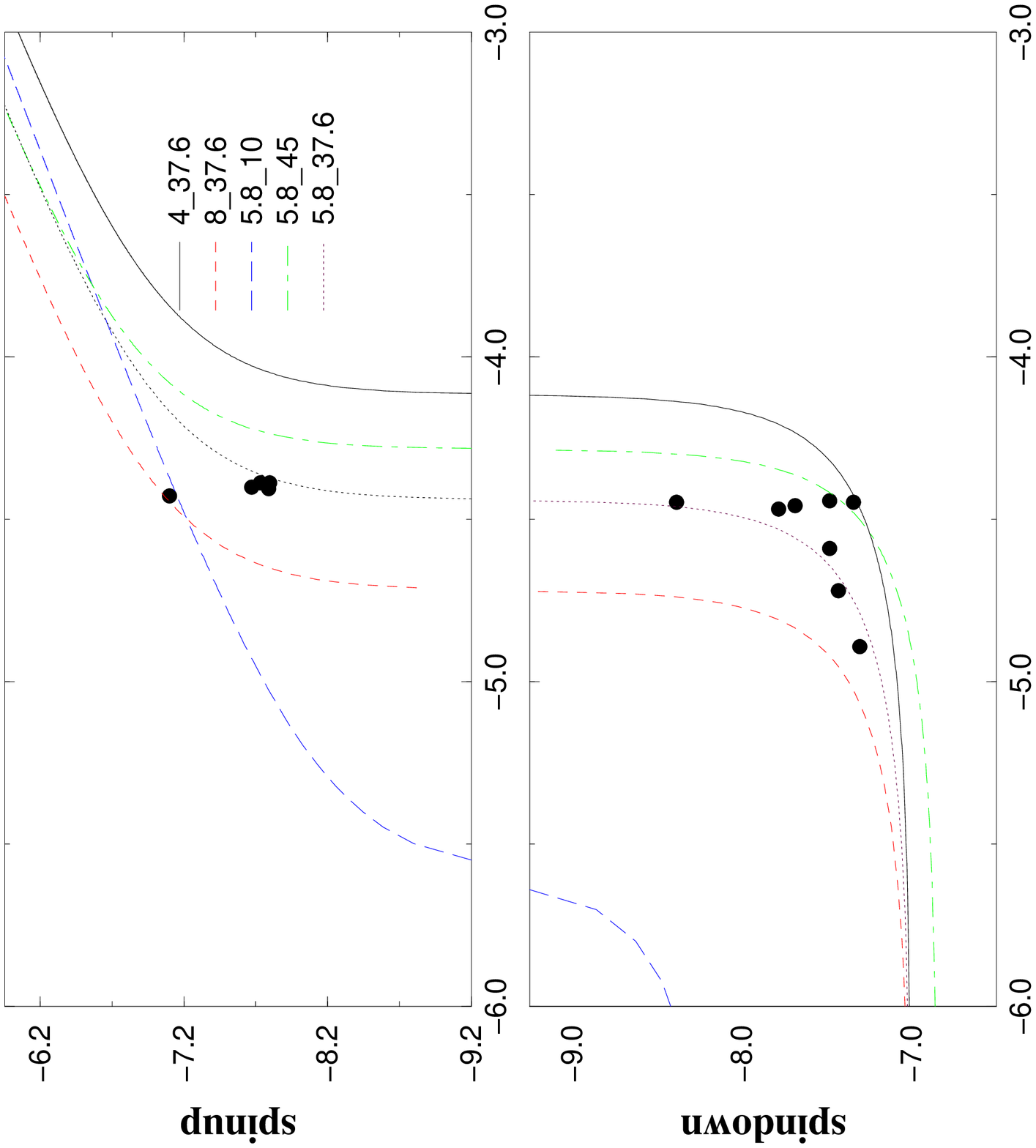}}}
\caption{Dependence of period derivative, $\dot p $, on the parameter $p^{7/3}f$, 
$f$-- observed flux, for GRO 1008-57. Both are axis in logarithmic scale.
Observations (Bildsten et al., 1997) are shown with black dots.
Five curves are plotted for disk accretion for different values of
distance to the pulsar and NS magnetic momentum. Solid curve: 
$d=4 \, {\rm kpc}$,
$\mu=10\cdot 10^{30}\, {\rm G}\cdot {\rm cm}^3$. 
Dashed curve: $d=8 \, {\rm kpc}$, 
$\mu=10\cdot 10^{30}\, {\rm G}\cdot {\rm cm}^3$.
Long dashed curve: $d=8 \, {\rm kpc}$, 
$\mu=45\cdot 10^{30}\, {\rm G}\cdot {\rm cm}^3$.
Dot-dashed curve (the best fit): $d=5.8 \, {\rm kpc}$, 
$\mu=37.6\cdot 10^{30}\,
{\rm G}\cdot {\rm cm}^3$. 
Dotted curve: $d=4 \, {\rm kpc}$, $\mu=45\cdot 10^{30}\, {\rm G}\cdot 
{\rm cm}^3$.}
\end{figure}

In more clear cases (Her X-1, GRO 1744-28), where we are sure, that
accretion is of the disk type, our estimates from maximum spin-down are in
good correspondence with observations. And we predict for the cases
of Be-transients, where disk accretion is working for sure, 
that in 2S 1417-624,
GRO 1948+32, GRO 1008-57, A 1118-616 and 4U 1145-61 
observations of cyclotron lines at
energies $\ge 100$ keV are possible in future.

 Observations of period and flux variations can be used also for
simultaneous determination of magnetic field of a NS and distance
to the X-ray source (Popov\cite{6}).

The method is based on several measurements
of period derivative, $\dot p$, and X-ray pulsar's flux, $f$.
Fitting distance, $d$, and magnetic momentum, $\mu$,
one can obtain good correspondence with the observed $p, \,\dot p$ and $f$,
and that way produce good estimates of distance and magnetic field
(see also another way of estimating of these parameters based
on the equilibrium period and spin-up measurements applied to GRO1744-28 in
(Joss \& Rappaport\cite{3}).

 Lets consider only disk accretion due to application of our method
to the system, in which most probably accretion is of the disk type. 
In that case one can write (see Lipunov\cite{4}):

\begin{equation}
\dot p= \frac {4 \pi ^2 \mu ^2}{3\,G\,I\,M} - \sqrt{0.45}\, 2^{-1/14}\frac
{\mu^{2/7}}{I} \left(GM\right)^{-3/7} \left[p^{7/3}L\right]^{6/7}R^{6/7},
\end{equation}

where $L=4\pi d^2 \cdot f$ -- luminosity, $f$ -- the observed flux.

So, with some small uncertainty 
in the equation above we know all parameters ($I$, $M$, $R$ etc.) 
except $\mu$ and $d$.
Fitting observed points with them we can obtain estimates of $\mu$ and $d$.
Uncertainties mainly depend on applicability of that simple model.

 To illustrate the method, we apply it to the X-ray pulsar 
GRO J1008-57, discovered by BATSE (Bildsten et al.\cite{1}). It is a 
$93.5 $ s X-ray pulsar, with the BATSE flux about $10^{-9}$ erg$\,$ cm$^{-2}$
s$^{-1}$. 
The source was
identified with a Be-system 
with $\sim 135^d$ orbital period.

 On figure 2 we show observations (as black dots)
and calculated curves for the disk model
on the plane $\dot p$ -- $p^{7/3} f$, where $f$ -- observed flux (logarithms
of these quantities are shown).
Curves were plotted for different values of the source distance, $d$,
and NS magnetic momentum, $\mu$. Spin-up and spin-down rates were obtained
from graphs in Bildsten et al.\cite{1}.

If one uses maximum spin-up, 
or maximum spin-down values to evaluate parameters of the pulsar,
then one can obtain values different from the best fit
(they are also shown on the figure): $d\approx 8 \,$ kpc, 
$\mu\approx 37.6\cdot 10^{30}\,$ G$\cdot$ cm$^3$
for maximum spin-up, and two values
for maximum
spin-down: $d\approx 4 \, {\rm kpc}$, 
$\mu\approx 37.6\cdot 10^{30}\,$ G$\cdot$ cm$^3$ and the one close to our
best fit (two similar values of maximum spin-down were observed
for different fluxes, but we mark, that formally maximum spin-down
corresponds to the values, which are close to our best fit). 
It can be used as an estimate of the errors of our method:
accuracy is about the factor of 2 in distance, and about the same value in
magnetic field, as can be seen from the figure.

Determination of magnetic field (and, probably, distance) only from X-ray
observations can be very useful in uncertain situations, for example, when
only X-ray observations
without precise localizations are available.

\noindent
{\bf Acknowledgments}

PSB thanks prof. Joss for discussions.
The work was supported by the RFBR (98-02-16801) and
the INTAS (96-0315) grants.

\end{document}